\documentclass[a4paper,11pt]{article}
 \usepackage{a4wide}

\usepackage{amsmath,amssymb,amsthm}
\usepackage{dsfont}
\usepackage{graphicx}
\usepackage[british]{babel}
\usepackage[utf8]{inputenc}
\usepackage{xcolor}
\usepackage[colorlinks=true]{hyperref}
\usepackage{authblk}
\usepackage{blindtext}
\usepackage{epstopdf}
\usepackage{cite}

\newtheorem{rem}{\bf Remark}[section]

\newcommand{\R}{\mathbb{R}}
\newcommand{\be}{\begin{equation}}
\newcommand{\ee}{\end{equation}}
\newcommand{\fer}[1]{(\ref{#1})}

\def\e{\epsilon}

\def\bN{{\bf N}}
\def\bP{{\bf P}}
\def\bQ{{\bf Q}}
\def\bbf{{\bf f}}
\def\bC{{\bf C}}
\def\mC{{\mathcal C}}

\begin{document}

\title{Optimal control of epidemic spreading in presence of social heterogeneity}

\author[1]{G. Dimarco\thanks{\tt giacomo.dimarco@unife.it}}
\author[2]{G. Toscani\thanks{\tt giuseppe.toscani@unipv.it}}
\author[3]{M. Zanella\thanks{\tt mattia.zanella@unipv.it}}

\affil[1]{\normalsize
        Department of Mathematics and Computer Science,
        University of Ferrara, Italy.} 
\affil[3]{
        Department of Mathematics "F. Casorati", 
        University of Pavia, Italy.}
\date{}

\maketitle

\begin{abstract}
	The spread of COVID-19 has been thwarted in most countries through non-pharmaceutical interventions. In particular, the most effective measures in this direction have been the stay-at-home and closure strategies of businesses and schools. However, population-wide lockdowns are far from being optimal carrying  heavy economic consequences. Therefore, there is nowadays a strong interest in designing more efficient restrictions. In this work, starting from a recent  kinetic-type model which takes into account the heterogeneity described by the social contact of individuals, we analyze the effects of introducing an optimal control strategy into the system, to limit selectively the mean number of contacts and reduce consequently the number of infected cases. Thanks to a data-driven approach, we show that this new mathematical model permits to assess the effects of the social limitations.  Finally, using the model introduced here and starting from the available data, we show the effectivity of the proposed selective measures to dampen the epidemic trends. \\[+.2cm]
  \end{abstract}

\tableofcontents

\section{Introduction}
Non pharmaceutical interventions (NPIs) had a strong impact in limiting the  spreading of the SARS-CoV-2 \cite{Ban}. The introduced NPIs span from basic personal hygiene precautions to more severe actions such as the mandatory stay-at-home and business and schools closures.

These measures were intended to diminish transmission rates and, to this end, aimed at reducing person-to-person contacts via social distancing. 
In fact, even if the precise role played by those interventions has still to be carefully understood in view of a deeper comprehension of the transmission mechanisms of the disease \cite{Ben,Bon}, it has however been recognized how the social attitudes of a population is a crucial feature for the spreading of airborne diseases \cite{APZ,BBT,Plos,DolT,Fuma}. Thus, it appears that understanding the interplay between social heterogeneity and clinical characteristics of a disease is an essential feature to design efficient and robust countermeasures in the near future. 

The  classical epidemiological models \cite{HWH00} are usually based on a compart\-men\-ta\-li\-za\-tion of the population, denoted in the following by $\mathcal C$, in which the agents composing the system are split into epidemiologically relevant states, like susceptible, who can contract the disease, and several other classes of agents who may transmit it. Therefore, to model the transmission of a pathogen through a population, if  $N_i(t)$ represents the number of individuals in the compartment $i\in \mathcal C$ with $|\mathcal C| = n>1$, one generally considers a system of differential equations that can be written in compact form as
 \be\label{classic}
 \frac {d\bN(t)}{dt} = \bP(\bN(t)),
 \ee
where $\bN(t)$ is the $n$-dimensional vector with components $N_i(t)$, and $\bP$ is the $n$-dimensional vector whose components $P_i(\bN)$, $i\in \mathcal C$, are the transition rates between compartments, namely the variation of the number of individuals in the $i$-th compartment due to interactions with individuals in other compartments. 

Several variations of system \fer{classic} have been introduced to forecast the penetration of a virus and to prevent its spread, see e.g. \cite{AM, BCF, DH,HWH00}. Generalizations of the mentioned compartmental models include additional factors characterizing the transmission dynamics, such as the introduction of age structure, contacts heterogeneity and mobility, see e.g. \cite{APZ,Arenas,CV,Cota}. 

In this work, we propose a kinetic approach to derive consistent transition rates in a population characterized by a heterogeneous number of contacts together with selective NPIs. Indeed, the number of social contacts has been recently identified by the scientific community as one of the main relevant causes of the potential pathogen transmission \cite{Plos,DolT,Fuma,Moss}. 
In this respect, the NPIs imposed by the different governments were mainly based on restrictions {whose} scope was to reduce as much as possible the number of contacts among individuals.

In more details,  we aim at giving a deeper understanding of the mitigation effects due to the reduction of the social interactions among individuals in the dynamics of SARS-CoV-2. This is done through a combination of a multiscale mathematical model which takes into account both the statistical information on the distribution of social contacts in a society as well as a control strategy whose target is to point the population towards a given benchmark number of contacts. The underlying theoretical framework that we consider is the one coming from the kinetic theory of collective social phenomena \cite{PT13}. The diffusion of epidemics is here described as a {result} of the interactions between a large number of individuals each one having a different social propensity. In this respect, as shown in \cite{DPTZ} and \cite{Zanella}, the fundamental tools of statistical physics are capable to provide a useful insight on epidemiological dynamics linking classical compartmentalization approaches with a statistical understanding of the social aspects. In fact, the essential multiscale nature of kinetic theory permits to determine the resulting macroscopic (or aggregate) and measurable characteristics of the evolution of the disease \cite{Zanella}. One of the key directions we pursue in this research is given by the combination of consolidated theoretical and modeling tools \cite{PT13} with data-driven techniques. More precisely, starting from the data at disposal, we calibrate our mathematical model passing through the determination of the relevant epidemiological parameters of the epidemic from the data. Subsequently, we concentrate ourselves on modeling the lockdown policies through an optimal control approach. In particular, we show how different kinds of control may result in very different mitigation effects, that deeply depend on the heterogeneity in the contact distribution of the population. In the last part, starting from the calibrated model at our disposal, we dedicate ourselves to the study of the effects induced by alternative control strategies.

We stress that, even if in the rest of the paper we mainly refer our results to a compartmental model with exposed and asymptomatic, i.e. the so-called SEIAR compartmentalization, the present approach is quite general, and applies both to simple \cite{AM, BCF, DH,HWH00} as well as to more complex epidemiological models \cite{Bruno,Gatto,LT}. For this reason we will keep the presentation as general as possible.

\section{Modeling epidemic dynamics in presence of contact heterogeneity}

We discuss in this part a possible improvement of the classical compartmental description of epidemic spreading which takes into account the statistical description of the social behavior of individuals. This permits to determine the distribution of the average number of connections in a system of $N$ interacting individuals, see \cite{Zanella,DPTZ}, and to relate this quantity with the infection dynamics. In this new approach, agents which are supposed to belong to a given compartment in term of response to a given pathogen, are additionally characterized by their daily number of contacts $x \ge 0$, whose statistical description is obtained through the distribution functions $f_i(x,t)$, $t \ge 0$, where $i \in \mathcal C$ denotes the $i$-th compartment (Susceptible, Infected, Asymptomatic, etc..), such that 
\[
\sum_{i \in \mathcal C}  f_i(x,t) = f(x,t), \qquad \int_{\mathbb R_+}f(x,t)dx = N(t). 
\]

In this setting, the time evolution of the functions $f_i(x,t)$, $i\in \mathcal C$ is obtained by integrating the compartmentalization epidemiological description with the simultaneous characterization of the formation of social contacts in a society \cite{DPTZ}. This merging is expressed by the system
\be\label{new}
 \frac {\partial \bbf(x, t)}{\partial t} = \bP(x, \bbf(x, t)) + \frac 1\tau \bQ(\bbf(x,t)).
\ee
As in \fer{classic}, $\bP$ is the $n$-dimensional vector whose components $P_i(x,\bbf)$, $i\in \mathcal C$, are the transition rates between compartments (now depending both on the contact variable $x$ and on the vector $\bbf(x,t)$), while $\bQ$ is the $n$-dimensional vector whose components $Q_i(f_i)$, $i\in \mathcal C$ are suitable relaxation operators which describe the formation of equilibrium distribution of contacts of the underlying compartments. Details about this emerging equilibrium distribution, together with its main properties, will be provided in the following. Last, $\tau>0$ measures the time-scale at which these equilibrium distributions of social contacts are reached by relaxation. 

In general, the description of the time evolution of epidemic spreading in terms of the distributions $f_i(x,t)$, as given by \fer{new}, is richer than the classic one obeying to \fer{classic}. This is due to the fact that, in \fer{new}, the system depends on an additional variable allowing a more realistic characterization of the transition rates between compartments. In particular, by assuming that the pathogen transmission does not depend upon the contact among individuals, the proposed description \fer{new} collapses to the standard approach \fer{classic} as shown in \cite{DPTZ}. 

In our model, the transition rates can now be assumed dependent on the intensity of social contacts in the compartments. As detailed later, in fact, they are obtained taking into account, in a very natural way, the heterogeneity of the population in the form of the variance of the distributions $f_i(x,t)$ \cite{DPTZ,Novo}. This point of view provides a better understanding of the role of the different individual behaviors in the spreading of a disease \cite{BBT}. 

In general, the knowledge of the densities $f_i(x,t)$ allows to evaluate various macroscopic quantities, independent from the number of contacts, the so-called moments of the distributions, obtained by integration with respect to the variable $x\in \mathbb R_+$. In this respect, the zero-order moment gives the number of individuals $N_i(t)$ in each compartment at time $t \ge 0$
\be\label{mass1}
N_i(t) = \int_{\R^+} f_i(x,t)\, dx ,\quad   i \in \mathcal C.
\ee
In the same way, one can quantify the mean numbers $\bar x_i(t)$ of contacts of individuals in each compartment at time $t \ge 0$ evaluating 
\be\label{mean1}
\bar x_i(t) = \frac 1{N_i(t)}\int_{\R^+}x\, f_i(x,t)\, dx, \quad   i \in \mathcal C.
\ee

Let now $f_S$, $S \in \mathcal C$, indicate the contact distribution of susceptible people (people at risk of becoming infected), and let the subset $\mathcal I\subset \mathcal C$, $|\mathcal I| < |\mathcal C|$ include the compartments of people that may transmit the infection. Since the transition rate $P_S(x, \bbf(x,t))$ contains the details about the spreading of the infection among susceptible individuals, the new formulation \fer{new} allows in particular to characterize this transmission in terms of the intensity of social contacts between the compartments of susceptible and infectious individuals. This intensity is measured at time $t \ge 0$ by relating the densities $f_S(x,t)$ and $f_j(x,t)$, $j \in \mathcal I$. Following \cite{DPTZ}, the transition rate $P_S$ is then expressed by introducing the function
 $K(f_S,\{f_j\}_{j \in \mathcal I})$, the so-called local incidence rate, given by
 \be\label{inci}
 K(f_S,\{f_j\}_{j \in \mathcal I})(x, t) = f_S(x,t) \int_{\R^+}  \sum_{j \in \mathcal I}\kappa_j(x,y)f_j(y,t) \,dy.
 \ee
In \fer{inci} the contact functions $\kappa_j(x,y)$ are nonnegative functions growing with respect to the number of contacts $x$ and $y$ of the populations of susceptible and infectious agents in the compartment $j \in \mathcal I$, and such that $\kappa_j(x, 0) = 0$.  In particular, the choice
\[
\kappa_j(x,y) = \beta_j\, x^\alpha y^\alpha,\qquad j \in \mathcal I
\]
with constants $\alpha, \beta_j >0$, corresponds to consider an incidence rate dependent on the product of the number of contacts of susceptible and infectious people. When $\alpha=1$, for example, the incidence rate takes the simpler form
\be\label{simple}
K(f_S,\{f_j\}_{j \in \mathcal I})(x,t) = x f_S(x,t) \sum_{j \in \mathcal I} \beta_j \bar x_j(t) N_j(t).
\ee

We discuss now the details of the dynamics of formation of the social contacts which are modeled by the operators $Q_i(f_i)$ in \eqref{new}. These operators are of Fokker-Planck-type with variable coefficient of diffusion and linear drift
\be\label{FPP}
Q_i (f_i(x,t)) = \frac 12 \frac{\partial}{\partial x}\left[\sigma^2  \frac{\partial(xf_i(x,t))}{\partial x} +  \mu \left( \frac x{\bar x_i(t)} -1 \right)f_i(x,t)\right], \qquad i \in  \mathcal C, 
\ee
coupled with no-flux boundary conditions such that the  Fokker-Planck-type equation
\[
\dfrac{\partial}{\partial t} f_i(x,t) = Q_i(f_i)(x,t)
\]
is mass and momentum preserving. In \fer{FPP}  $\sigma$ and $\mu$ are positive constants which characterize the intensities of the diffusion and drift terms. The above description through Fokker-Planck operators is obtained thanks to an upscaling of an underlying microscopic dynamics which takes into account the typical aspects of human behaviour for what concerns the social attitudes \cite{DPTZ}. In particular, the microscopic description takes into account the interactions due to the basic daily needs such as work duties, schools and the search, in absence of epidemics, of opportunities for socialization. These aspects are embedded in the microscopic model by relying on the so-called prospect theory of Kahneman and Tversky \cite{KT1} as detailed later.

The kernels of the operators $Q_i$ furnish the class of equilibrium densities, that are obtained by solving the first-order differential equations
 \[
 \sigma  \frac{\partial (xf_i^\infty(x))}{\partial x} +  \mu \left( \frac x{\bar x_i} -1 \right)f_i^\infty(x)= 0, \quad i \in \mathcal C, \quad \sum_{i\in C}N_i(t)=N(t).
 \]
 It is immediate to verify that, in this setting, the equilibrium distributions $f^\infty_i(x)$ of given mass $N_i$ and local mean $\bar x_i$ are the Gamma distributions
  \be\label{eq1}
f^\infty_i(x) = f^\infty_i(x, \bar x_i, \lambda) =  N_i \left( \frac \lambda{\bar x_i}\right)^\lambda \frac 1{\Gamma(\lambda)} x^{\lambda -1} \exp\left\{ - \frac \lambda{\bar x_i} x\right\},
 \ee 
 characterized by  the shape parameter $\lambda = \mu/\sigma$. It is important to stress the role of
 the positive constant $\lambda$ which is a measure of the heterogeneity of the population, statistically distributed according to \fer{eq1}. It is also remarkable to notice that the above equilibrium dynamics, satisfying the condition $\lambda>1$, is in agreement with the experimental results in \cite{Plos}. This study represents one of the first attempts in which a large scale experimental measure of the relation between contact and spread of infectious diseases has been done. To better characterize the equilibrium distribution, we also notice  that the condition on $\lambda$ identifies the Gamma densities for which $f^\infty_i (x=0) = 0$. Moreover,  higher order moments of this class of distributions depend only on  $N_i$, $\bar x_i$ and $\lambda$. In particular, the local variance is expressed by
\be\label{vari}
V_i =  \frac 1{N_i}\int_{\R^+}\left( x -\bar x_i\right)^2 f^\infty_i(x,\bar x_i, \lambda)\, dx = \frac {\bar x_i^2}{\lambda}.
\ee
Hence,  expression \fer{vari} permits to relate the large values of $\lambda$ with small heterogeneity of the population as claimed before.

Before introducing in the model the aspects related to the control policies, we briefly discuss how the Fokker-Planck-type operators \eqref{FPP} can be derived from collisional kinetic equations describing the evolution of social connections \cite{GT19,To3}. These kinetic equations are obtained by looking to a possible characterization of the personal behavior of a single agent in terms of daily contacts starting from some well-established assumptions. Indeed, the total amount of contacts can be simply view as the result of a repeated upgrading: the daily life of each person is based on a certain number of activities each carrying a certain number of contacts. Moreover, independently of the personal choices or needs, this number contains a certain amount of randomness, that has to be taken into account. These assumptions give, letting $x^\prime$ be the microscopic number of social contacts resulting from the interaction,
\be\label{coll1}
x^\prime = x - \Phi_\epsilon(x/\bar x_{i})x + \eta_\epsilon x, 
\ee
with $\eta_\epsilon$ a centered random variable such that $\mathbb E[ \eta_\epsilon^2 ] = \epsilon \sigma^2$ with bounded moments up to order three,  and $\Phi_\epsilon(\cdot)$ a transition function of the form
\be\label{valuef}
\Phi_\epsilon(s) = \mu \dfrac{e^{\epsilon(s-1)}-1}{e^{\epsilon (s-1)}+1}, \qquad s\ge 0. 
\ee
The transition function has been constructed by taking into account that while is very easy to reach a high number of social contacts for need or will, going below a given threshold is very difficult, since contacts are forced by basic activities which cannot be avoided in general. This asymmetry between growth and decrease, as exhaustively discussed in \cite{DT,GT19,DPTZ,To3,To4}, can be suitably modeled by \eqref{valuef} which plays the role of the so-called \emph{value function} in the prospect theory of Kahneman and {Tversky} \cite{KT1} and implements the advocated asymmetry in the contact distribution formation.

Hence, with the above discussed choices, at the aggregate scale the evolution of a contact distribution $f_{i,\epsilon}$ is given by the Boltzmann-type equation in weak form 
\be
\label{eq:boltz}
\dfrac{\partial}{\partial t}\int_{\R_+}\varphi(x)f_{i,\epsilon}(x)dx = \dfrac{1}{\epsilon}\mathbb E\left[\int_{\mathbb R_+} B(x)\left( \varphi(x^\prime)- \varphi(x)\right)f_{i,\epsilon}(x,t) dx \right],
\ee
where $B(x) = 1/x$ is the interaction kernel assigning a low probability to interactions to individuals exhibiting a large number of contacts and assigning a high probability to interactions when $x$ is small. Letting finally $\epsilon \rightarrow 0^+$ (the limit of \emph{grazing} interactions) one shows that \eqref{eq:boltz} converges, up to extraction of a subsequence, to the Fokker-Planck equation \eqref{FPP}, see e.g. \cite{DPTZ,FPTT}. This result is particularly useful since it permits to characterize the equilibrium state of the system, i.e. it permits to find the equilibrium Gamma distribution \eqref{eq1} as discussed in \cite{DPTZ} and experimentally found in \cite{Plos}. Moreover, it allows to analytically consider and study control problems at the kinetic level, i.e. to define effective actions at the microscopic level which scope is to modify the behavior of the single individuals. Subsequently, thanks to well established kinetic theory tools \cite{PT13}, it is possible to upscale the microscopic actions in order to compute its reflection at the aggregate, measurable, scale of description and of interest. 

\subsection{Observable effects of selective social restrictions}
In this part we model the lockdown measures intended to reduce the social interactions by introducing an additive term $u$ at the level of the microscopic formation of the contacts \fer{coll1} to limit selectively the social activities. This gives the controlled elementary interaction 
\begin{equation}\label{uni}
x^\prime  = x -  \Phi_\e \left(\frac x{\bar x_{i}}\right) x + \sqrt{\e\tau}  W(x)\,u  + \eta_\e\, x,\qquad W\ge 0.
\end{equation}
The function $W(x)$ is such that if $W\equiv 1$ the control variable $u$ is independent from the number $x\ge 0$ of the social contacts, while if $W(x)\propto x$ the control is considered heavier for people with a higher number of contacts. Note that in \eqref{uni}, the size of the controlled variable is tuned by the small parameters $\e$ and $\tau$, which ensure that the control acts at the right scale with respect to both the grazing variable $\e$ characterizing the passage from a Boltzmann to a Fokker-Planck dynamics and the relaxation variable $\tau$ characterizing the speed at which an equilibrium state for what concerns the distribution of contact is reached, compared to the speed at which the epidemic dynamics travels. 

Then, the so-called optimal control $u^*$ is determined as the minimizer of a given cost functional 
\begin{equation}\label{eq:optimal_u}
u^* = \textrm{arg}\min_{u\in \mathcal U}  \mathcal J(x^\prime, u),
\end{equation}
subject to the constraint \eqref{uni}. In \fer{eq:optimal_u} the minimum is taken on the set $\mathcal U$ of all admissible controls. A classical choice for $\mathcal J$ is a quadratic cost functional of the form
\be\label{quadra}
\mathcal J_i(x^\prime , u)= \dfrac{1}{2}\mathbb E \left[ (x^\prime - x_{T,i})^2 + \nu\, u_i^2 \right] \qquad i \in \mathcal C,
\ee
being $\nu >0$ a penalization coefficient measuring the economic and social cost of the control and $x_{T,i}>0$ the desired target number of social contacts for the compartment $i \in \mathcal C$. According to \eqref{quadra}, the cost of the control increases quadratically with the distance to the desired target of average number of contacts. The presence in \fer{quadra} of the mean operator  is related to the fact that, by hypothesis, a control cannot act on the random fluctuations introduced to mimic the unanticipated daily interactions among agents. Let us observe that other convex cost functionals may be considered as well. However, they often lead to problems whose analytical solution cannot be obtained explicitly. 
Proceeding now as in \cite{PTZ} the control introduced in \fer{uni} can be solved explicitly giving
\be\label{eq:u_explicit}
u = -\dfrac{\sqrt{\epsilon \tau}W(x)}{\nu + \epsilon \tau W^2(x)}\left(x-x_{T,i}-\Phi_\epsilon(x/\bar x_{i})x\right), 
\ee
which generates the optimal constrained law of growth of the social contacts 
\begin{equation}\label{uni-c}
x_*^\prime  = x - \frac{\nu}{\nu + \e\tau W^2(x)} \Phi_\e \left(\frac x{\bar x_{i}}\right) x- \frac{\e\tau }{\nu + \e\tau W^2(x)}(x-x_{T,i})W^2(x) + \eta_\e\, x . 
\end{equation}

Now, in presence of the above described microscopic controlled interactions, one can consider as before the construction of a kinetic collisional equation for what concerns the description of the formation of the social structure
\be
\label{eq:boltz_cont}
\dfrac{\partial}{\partial t}\int_{\R_+}\varphi(x)f_{i,\epsilon}(x)dx = \dfrac{1}{\epsilon}\mathbb E\left[\int_{\mathbb R_+} \frac{\left( \varphi(x_*^\prime)- \varphi(x)\right)}{x}f_{i,\epsilon}(x,t) dx \right],
\ee
where we recall that the frequency of interaction has been chosen such that $B(x)=1/x$. Then, thanks to the passage to the \textit{grazing limit} one can show that the controlled kinetic model converges to a Fokker--Planck type equation which contains an additional drift term, that quantifies the role of the control mimicking lockdown measures. In fact, since by hypothesis $\epsilon\ll 1$, we can write
\[\varphi(x_*^\prime)-\varphi(x)=(x_*^\prime-x)\partial_x \varphi(x)+\frac{1}{2}(x_*^\prime-x)^2\partial_x^2 \varphi(x)+\frac{1}{6}(x_*^\prime-x)^3\partial_x^3 \varphi(\hat x)
\]
with $\hat x\in(\min(x,x_*^\prime),\max(x,x_*^\prime))$. Then using \eqref{uni-c}
one gets 
\begin{equation} 
\label{eq:boltz_cont1}
\begin{split}
&\dfrac{\partial}{\partial t}\int_{\R_+}\varphi(x)f_{i,\epsilon}(x)dx =  -\dfrac{1}{\epsilon}\int_{\mathbb R_+}\frac{\nu}{\nu + \e\tau W^2(x)}\Phi_\e(x/\bar x_i) \partial_x \varphi(x)f_{i,\epsilon}(x,t) dx\\
&-\int_{\mathbb R_+}\frac{\tau }{\nu + \e\tau W^2(x)}\frac{(x-x_{T,i})}{x}W^2(x)\partial_x \varphi(x)f_{i,\epsilon}(x,t) dx \\
&+\frac{\sigma^2}{2}\int_{\mathbb R_+}\partial_x^2 \varphi(x)x f_{i,\epsilon}(x,t) dx +R_{\Phi_\epsilon}(f_{i,\epsilon})(x,t),
\end{split}
\end{equation}
where the remainder $R_{\Phi_\epsilon}(f_{i,\epsilon})(x,t)$ can be shown to go to zero in the limit $\epsilon\to 0^+$ since by assumption $\varphi$ and its derivatives are bounded in $\R_+$ and the stochastic variable $\eta$ is assumed to have bounded moments at least of order 3 as previously stated. We point the reader to \cite{DPTZ,PT13} for a detailed discussion on this aspect. In the limit $\epsilon\to 0^+$ one gets
\begin{equation}
\begin{split}
	\label{eq:boltz_cont2}
	&\dfrac{\partial}{\partial t}\int_{\R_+}\varphi(x)f_{i}(x)dx =\\
	& -\int_{\mathbb R_+}\Phi(x/\bar x_i) \partial_x \varphi(x)f_{i}(x,t) dx
	-\int_{\mathbb R_+}\frac{\tau }{\nu }\frac{(x-x_{T,i})}{x}W^2(x)\partial_x \varphi(x)f_{i}(x,t) dx\\
	&+\frac{\sigma^2}{2}\int_{\mathbb R_+}\partial_x^2 \varphi(x)x f_{i}(x,t) dx
\end{split}
\end{equation}
since for $\epsilon\ll 1$ we have
$$\Phi_\epsilon(x/\bar x_i)\approx \epsilon\frac{\mu}{2}\left(\frac{x}{\bar x_i}-1\right) = \epsilon \Phi(x/ \bar x_i).$$
Integrating back by part \eqref{eq:boltz_cont2}, and considering the no flux boundary condition at $x=0^+$ to preserve mass and momentum,   the relaxation equations take the form
\be\label{FPC}
\frac{\partial f_i(x,t)}{\partial t} = \frac 1\tau Q_i(f_i(x,t)) + C_i(f_i(x,t)), \qquad i \in \mathcal C. 
\ee
where the operator $C_i(f_i)$ in \fer{FPC} is given by
 \be\label{con}
 C_i(f_i(x,t)) = \frac 1\nu \frac{\partial}{\partial x} \left[  \dfrac{x - x_{T,i} }{x} W^2(x)f_i(x,t) \right],
 \ee
 and $Q_i(f_i(x,t))$ remains the same as in \eqref{FPP}. Note in particular that in \fer{con} the control term does not depend on the time scale $\tau$ of the Fokker-Planck operator. 

Therefore, in presence of social restrictions modeled by the control term introduced above,  system \eqref{new} is modified as
\be\label{new-cont}
 \frac {\partial \bbf(x, t)}{\partial t} = \bP(x,\bbf(x, t)) +  \bC(\bbf(x,t)) + \frac 1\tau \bQ(\bbf(x,t)),
\ee
where the $n$-dimensional vector $\bC$ quantifies the effect of the control on the epidemic spreading. According to the previous analysis, the components of the vector control $\bC$ are given by the operators of type \fer{con}, where the values of the penalization coefficients and of the target values may depend on the specific compartment $i \in \mathcal C$. 

In the next part, we focus on a specific compartmental model for which we give the details of the controlled strategy for the reduction of the social contacts.

\begin{rem}
We observe that the cost functional \eqref{quadra} introduced at the level of microscopic interactions can be seen as an instantaneous approximation of a  macroscopic functional obtained by considering \eqref{eq:u_explicit} in the limit $\epsilon, \tau \rightarrow 0^+$ whose form is 
\[
J_i = \dfrac{1}{2}  \int_{\mathbb R_+} \left(1 + \dfrac{W^2(x)}{\nu}\right)(x-x_{T,i})^2  f_i^\infty(x)dx, \qquad i \in \mathcal C. 
\]
In the above functional we consider the contributions at the final time $T$ and the steady distribution $f_i^\infty$ of contacts defined in \eqref{eq1} instead of the instantaneous single interaction among agents.
\end{rem}
 
 \subsection{A SEIAR-type compartmentalization}
 We consider the case $\mathcal C = \{S,E,I,A,R\}$, i.e. we concentrate on a model with exposed and asymptomatic compartments and  we discuss the observable effects of the introduced control at the level of the mean number of connections in two leading examples. 

The choice $\mathcal C = \{S,E,I,A,R\}$ gives, accordingly to \eqref{new-cont}, the following system of kinetic equations relating the epidemiological dynamics with the number of daily contacts expressed in terms of a probability distributions $f_i(x,t)$
\be
\label{FPC2}
\begin{split}
\dfrac{\partial}{\partial t}f_S(x,t) &= - x f_S(x,t) \sum_{j\in\{I,A\}} \beta_j \bar x_j N_j(t) +C_S(f_S)(x,t)+ \dfrac{1}{\tau}Q_S(f_S)(x,t)\\
\dfrac{\partial}{\partial t}f_E(x,t) &= x f_S(x,t) \sum_{j\in\{I,A\}} \beta_j \bar x_j N_j(t) - \gamma_E f_E(x,t) +C_E(f_E)(x,t)  + \dfrac{1}{\tau}Q_E(f_E)(x,t) \\
\dfrac{\partial}{\partial t}f_I(x,t) &= \xi \gamma_E f_E(x,t) - \gamma_I f_I(x,t)+C_I(f_I)(x,t) + \dfrac{1}{\tau}Q_I(f_I)(x,t)\\
\dfrac{\partial}{\partial t}f_A(x,t) &= (1-\xi)\gamma_E f_E(x,t) - \gamma_A f_A(x,t)+C_A(f_A)(x,t) + \dfrac{1}{\tau}Q_A(f_A)(x,t) \\
\dfrac{\partial}{\partial t}f_R(x,t) &= \gamma_I f_I(x,t) + \gamma_A f_A(x,t)+C_R(f_R)(x,t) + \dfrac{1}{\tau}Q_R(f_R)(x,t)
\end{split}
\ee
where, to have a completely definite system, it remains to clarify the role of the parameter $\xi$ which gives the fraction of exposed individuals becoming successively infected and its counterpart $(1-\xi)$ which gives the fraction of asymptomatic individuals after the exposure to the virus. In addition, the functions $\gamma_I$ and $\gamma_A$ represent the recovery rates of infected and asymptomatic individuals respectively, which give the frequency at which individuals move to the  compartment of recovered individuals. In the present setting, the choice of the constants $\beta_I,\beta_A>0$ takes into account that the population of the infected compartment is considered recognized by the health system through a swab test and therefore subject to isolation. We collect in the asymptomatic compartment unrecognized infectious that are not subject to isolation. We highlight that we preferred to neglect the compartment of deceased people since they do not influence the spreading of the disease. Anyway, the inclusion of this new compartment to assess the risks related to the infection can easily be considered.

Since the $Q_i$ and $C_i$ components in \eqref{FPC2}, coupled with no-flux boundary conditions, are mass  preserving, we can integrate the equations of the system with respect to the contact variable $x$ to obtain a macroscopic system of equations describing the evolution of the fraction of individuals in each compartment independently on the number of daily social contacts. This gives the system
\be\label{seiar-mass}
\begin{split}
\dfrac{dN_S}{dt}  &= - \bar x_S N_S \sum_{j\in\{I,A\}} \beta_j \bar x_j N_j,\\
\dfrac{dN_E}{dt}  &= \bar x_S N_S \sum_{j\in\{I,A\}} \beta_j \bar x_j N_j - \gamma_E N_E, \\
\dfrac{dN_I }{dt} &= \xi \gamma_E  N_E - \gamma_I N_I \\
\dfrac{dN_A}{dt}  &= (1-\xi) \gamma_E N_E - \gamma_A N_A \\
\dfrac{dN_R}{dt}  &=  \gamma_I N_I + \gamma_A N_A.
\end{split}
\ee
which is not closed since the evolution in time of the vector solution $\bf N(t)$ depends on the mean number of contacts $\bar x_i$. Since the $Q_i$ components, under the same hypothesis on the boundary fluxes, are also momentum preserving, by multiplying the distributions $f_i(x,t)$ by $x$ and integrating with respect to the contact variable, it is  possible to recover a further macroscopic set of equations describing the constrained dynamics of the mean number of social contacts in each compartment. However, one immediately realizes that, this second set of macroscopic observables depends upon higher order moments of the distributions $f_i(x,t)$. In other words, in general, the system of macroscopic equations which can be obtained by integration of \fer{FPC2} over the number of social contacts is not closed. In fact, system \fer{seiar-mass} depends on the time evolution of $\bar x_i(t), \ i=S,E,I,A,R$ while the system giving the time evolution of $\bar x_i(t), \ i=S,E,I,A,R$ depend upon second order moments of the $f_i(x,t) , \ i=S,E,I,A,R$. This represents nothing else than the classical so-called closure problem often encountered in kinetic theory \cite{PT13}. 

However, letting $\tau \to 0$ forces the distributions $f_i(x,t)$ to converge towards the Gamma vector distribution of components \fer{eq1} with mass fractions $N_i(t)$, and local mean values $\bar x_i(t)$, $i \in \mathcal C = \{S,E,I,A,R\}$. This amounts to assuming that social dynamics are much faster than epidemic dynamics. In this situation, it is possible to obtain a closed set of macroscopic equations. In particular, the second order moments of the distributions $f_i(x,t)$ can be explicitly computed in terms of the mass fractions and mean values resorting to \fer{vari}. This gives the system
\be\label{seiar-mean}
\begin{split}
\dfrac{d \bar x_S}{dt}  &= -  \dfrac{1}{\lambda}  \bar x_S^2  \sum_{j \in \{I,A\}} \beta_j \bar x_j N_j + \mC(f_S^\infty) \\
\dfrac{d \bar x_E}{dt}  &= - \bar x_S\frac{N_S}{N_E}\left( \dfrac{\lambda+1}{\lambda}  \bar x_S - \bar x_E \right) \sum_{j \in \{I,A\}} \beta_j \bar x_j N_j + \mC(f_E^\infty)  \\
\dfrac{d \bar x_I}{dt}  &= \xi \gamma_E\left( \bar x_E - \bar x_I\right)\frac{N_E}{N_I}  + \mC(f_I^\infty) \\
\dfrac{d \bar x_A}{dt}  &= (1-\xi)\gamma_E \left( \bar x_E - \bar x_A\right)\frac{N_E}{N_A}  + \mC(f_A^\infty) \\
\dfrac{d \bar x_A}{dt} &= \gamma_I \left( \bar x_I - \bar x_R\right)\frac{N_I}{N_R}  + \gamma_A \left( \bar x_A - \bar x_R\right)\frac{N_A}{N_R} + \mC(f_R^\infty).
\end{split}
\ee
where
\be\label{cont-gen}
\mC(f_i^\infty)(t) = \dfrac{1}{\nu\, N_i(t)}  \int_{\R_+} \left(\dfrac{x_{T,i}}{x}-1 \right) W^2(x)f^\infty_i(x,t)dx. 
\ee
Equations \fer{seiar-mean} coupled with the system \fer{seiar-mass} describe in closed form the time evolution of an epidemic where the transition functions from one compartment to another depend upon the mean level of social contacts in the population. In addition, these latter depend on the heterogeneity of the population through the parameter $\lambda=\mu/\sigma$ characterizing the distributions $f_i(x,t)$, while the term \fer{cont-gen} models the effects of the lockdown measures.
 
In the rest of the paper, we focus on two specific benchmark cases for which it is possible to compute explicitly the effect of the social restrictions on the system. In particular, we first concentrate on the case of uniform restrictions, corresponding to $W(x)\equiv 1$, and second on the case of selective restrictions. This last situation describes an agent system  in which restrictions are stronger for individuals exhibiting a larger number of contacts,  and it is based on the choice  $W(x) = x/\bar x_i$. By direct computation we have 
\be
\label{eq:Cinf}
\mC (f_i^\infty) = 
\begin{cases}
\dfrac{1}{\nu}\left(\dfrac{\lambda}{\lambda-1} \dfrac{x_{T,i}}{\bar x_i} - 1 \right) & W(x) = 1 \\
\dfrac{1}{\nu} \left( \dfrac{1+\lambda}{\lambda}\bar x_i - x_{T,i} \right) & W(x) = \dfrac{x}{\bar{x}_i}
\end{cases}
\ee

Hence, for small penalizations $\nu>0$ of the control, the mean number of connections stabilizes towards the values
\[
\bar x_i^\infty = 
\begin{cases}
x_{T,i} \dfrac{\lambda}{\lambda-1}>x_{T,i} & W(x) = 1 \\
 \dfrac\lambda{1+\lambda} x_{T,i} < x_{T,i} & W(x) = \dfrac{x}{\bar{x}_i}.
\end{cases}
\]
It is worth to notice that, in both situations, the role of the penalization parameter $\nu>0$ is limited to a scaling of time, without changing the final goal of the control. Thus, a smaller penalization parameter is reflected in a quicker convergence of the solution towards the target value as expected. We also notice that the selective control, i.e. the one especially operating on individuals with larger social nets, allows to reach for the global population a value below the desired target $x_T$. In particular, the mean contact rate is lower than $x_T$ of an amount directly proportional to the degree of heterogeneity of the population. This can be explained by the fact that individuals with larger social networks are the main responsible for the creation of the contact dynamics. This situation suggests that it is important to act on this class of people in order to limit the spread of a disease instead of taking measures with uniform impact over the whole population.

From an operative point of view, it is clear that the application of a uniform control, which is based on restrictive measures uniformly valid for the global population, are simpler to apply with respect to measures which provide for heavier restrictions for the categories that show a greater number of social contacts. In this direction, the previous analysis  seems to suggest that a mixture of the two types of controls will present the best compromise between containment results and effective applicability of containment measures. 

\section{Numerical tests}

In this section using \eqref{seiar-mass}-\eqref{seiar-mean}, we compare possible interventions for the containment of an epidemic based on the limitation of the mean number of contacts. These non-pharmaceutical measures are designed to drive the population towards a prescribed target dependent on the relevant epidemiological state of agents, namely $x_{T,i}$, $i \in \mathcal C$ in the considered framework. 

To this end, we first estimate the relevant epidemiological parameters of the second order macroscopic SEIAR model taking into account the available dataset provided by John Hopkins University. \footnote{\texttt{https://github.com/CSSEGISandData/COVID-19}} After the calibration of the model with the data at disposal, we assess the impact of different control strategies leading to the observed epidemic dynamics in different countries where lockdown policies have been implemented.  

We will mainly focus on the so-called first wave of infection lasting the first half of 2020 since in this period similar contact restrictions have been proposed in all Western countries. In the following we concentrate on the cases of Italy and United Kingdom.  

\subsection{Calibration of the second order model}

\begin{table}\centering
\caption{Timeline of COVID-19 pandemic}
\label{tab:dates}
\begin{tabular}{ccccc}
\hline
 Country & First detected case  & Begin restrictions & $\beta_I$ & $\lambda$  \\
\hline
Italy                    & January 30th         & March 9th & 0.0042 & 4.9593\\
United Kingdom & January 31st         & March 23rd & 0.0043 & 5.8502 
\end{tabular}
\vspace*{-4pt}
\end{table}

We consider two countries that adopted the stay-at-home policies during early stages of the infection. In Table \ref{tab:dates} we report the days of implementation of such measures together with a sketch of the timeline of the epidemic. In the second column, we highlight the begin of restrictions at the global level for each different country. Anyway, it is worth to mention that often local restriction measures, i.e. isolation of a small portion of a territory, have been implemented  as well. 

\begin{figure}
\centering
\includegraphics[scale = 0.33]{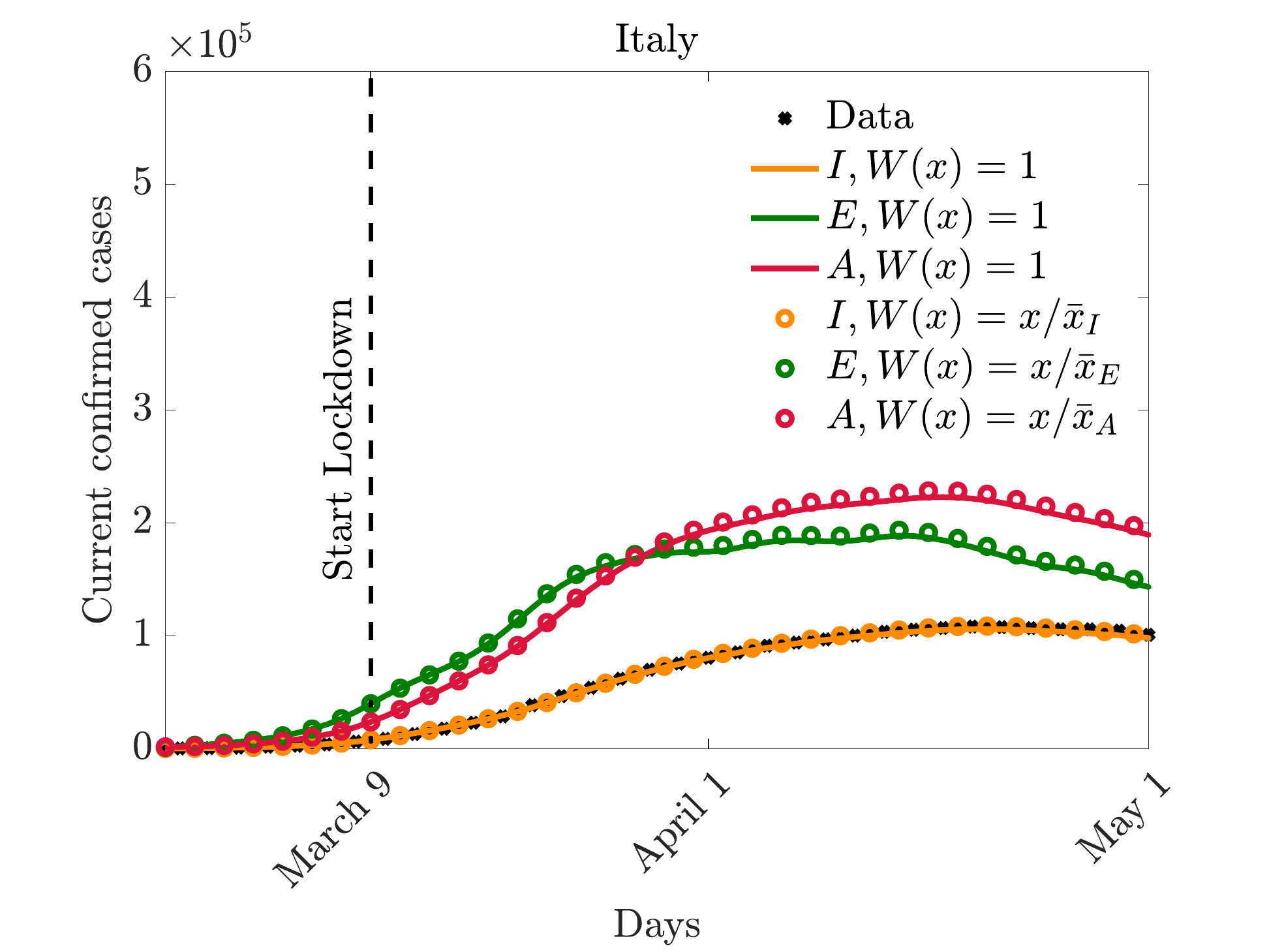}
\includegraphics[scale = 0.33]{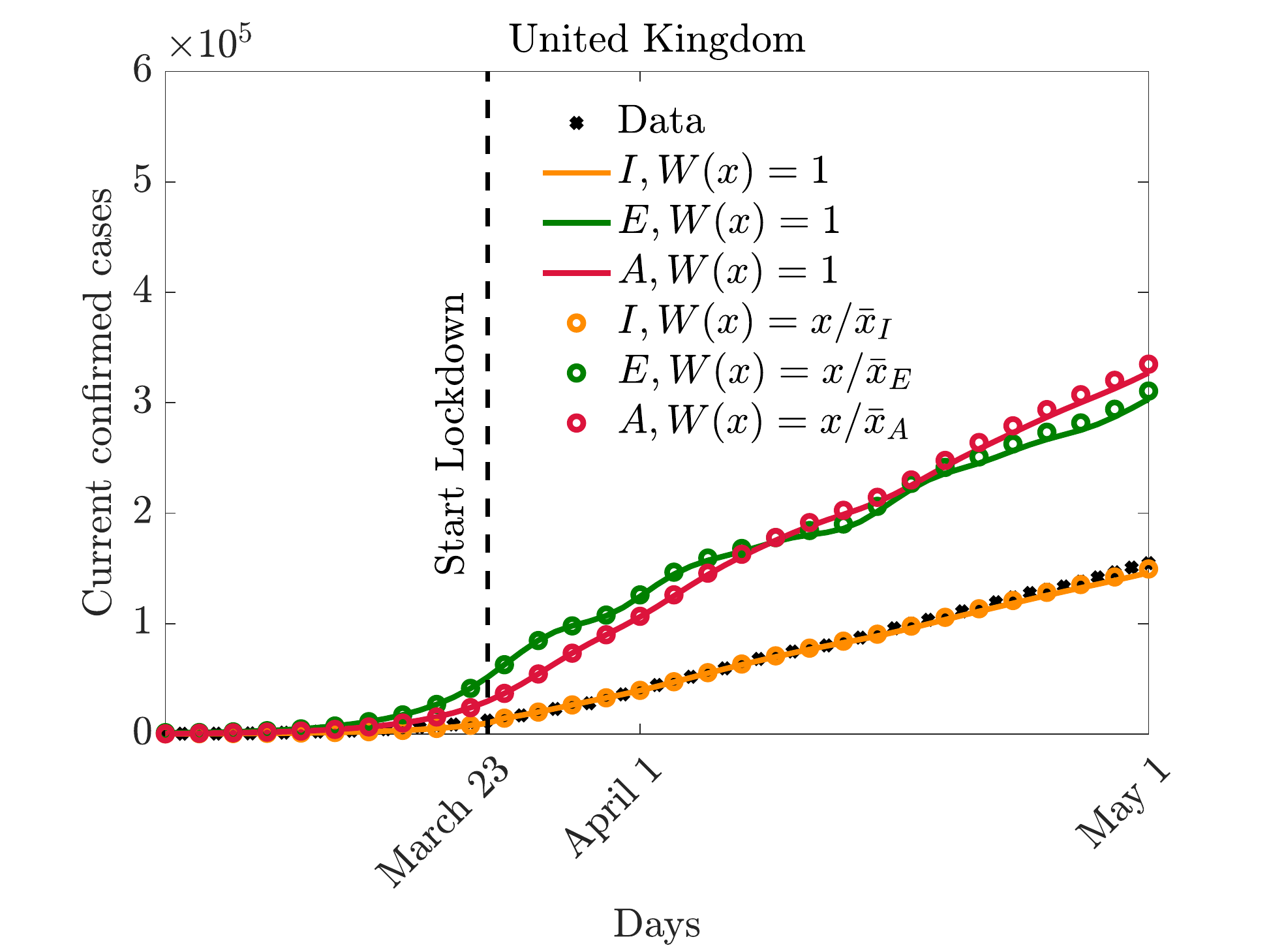}\\
\includegraphics[scale = 0.33]{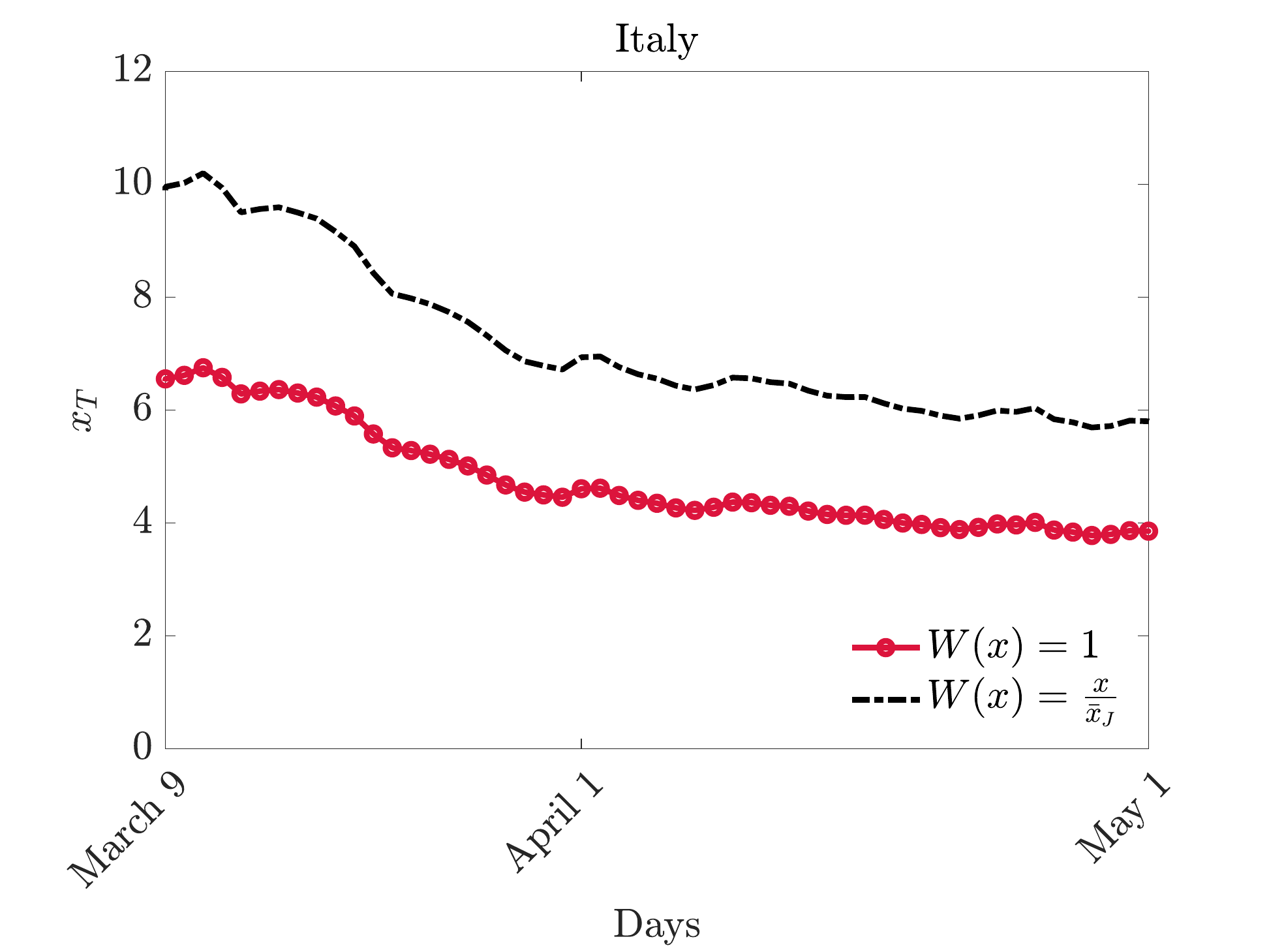}
\includegraphics[scale = 0.33]{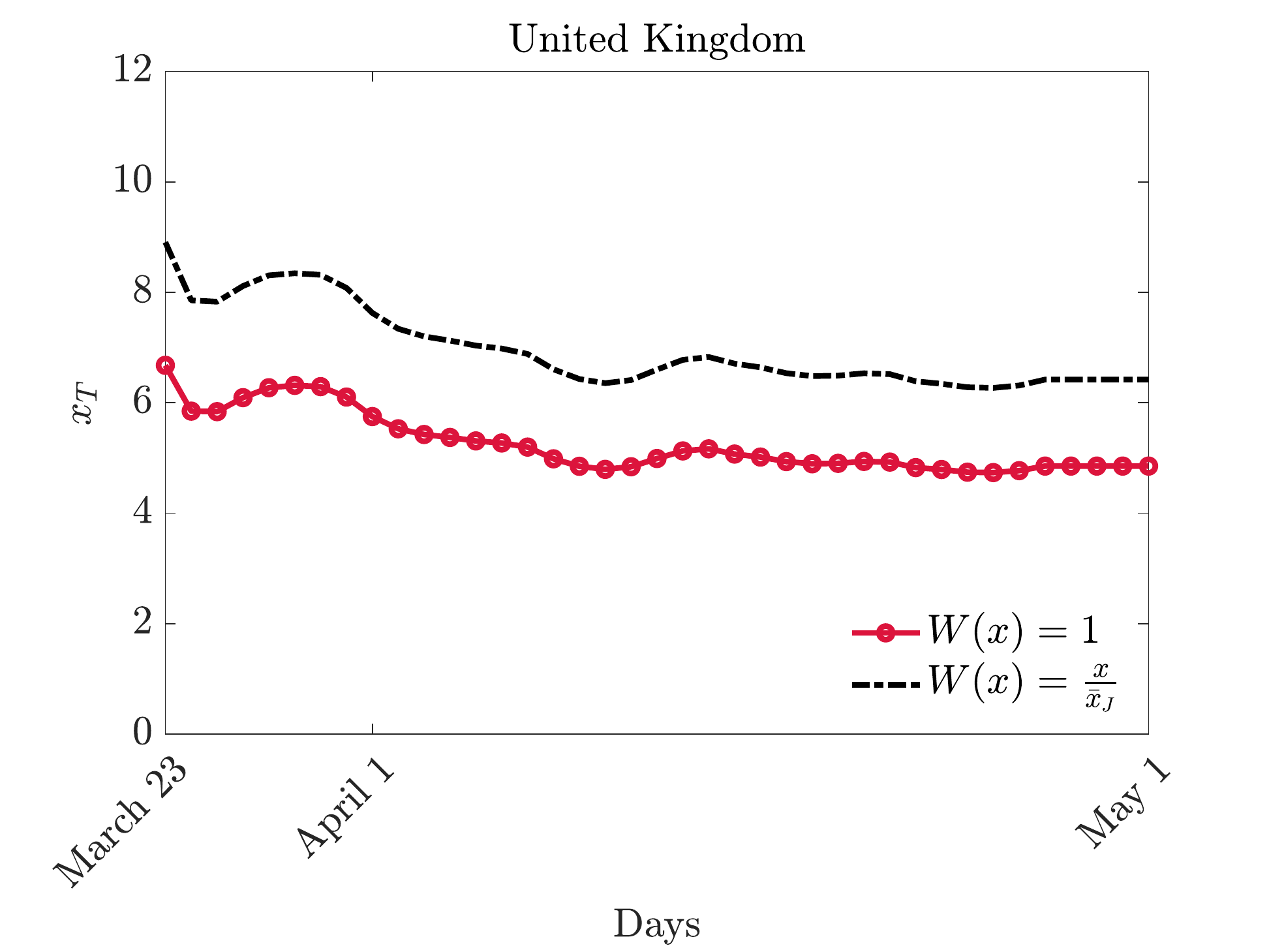}\\
\includegraphics[scale = 0.33]{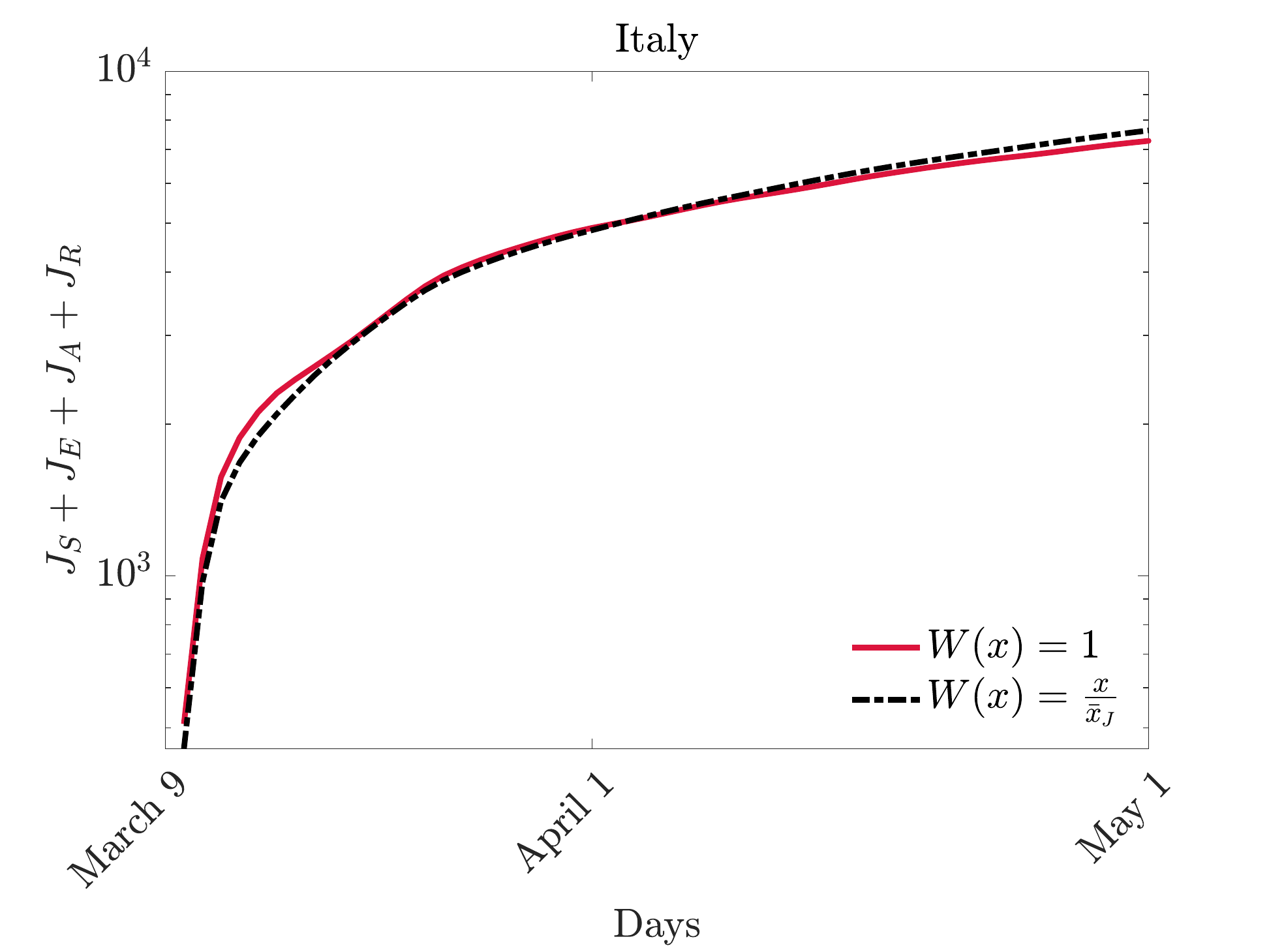}
\includegraphics[scale = 0.33]{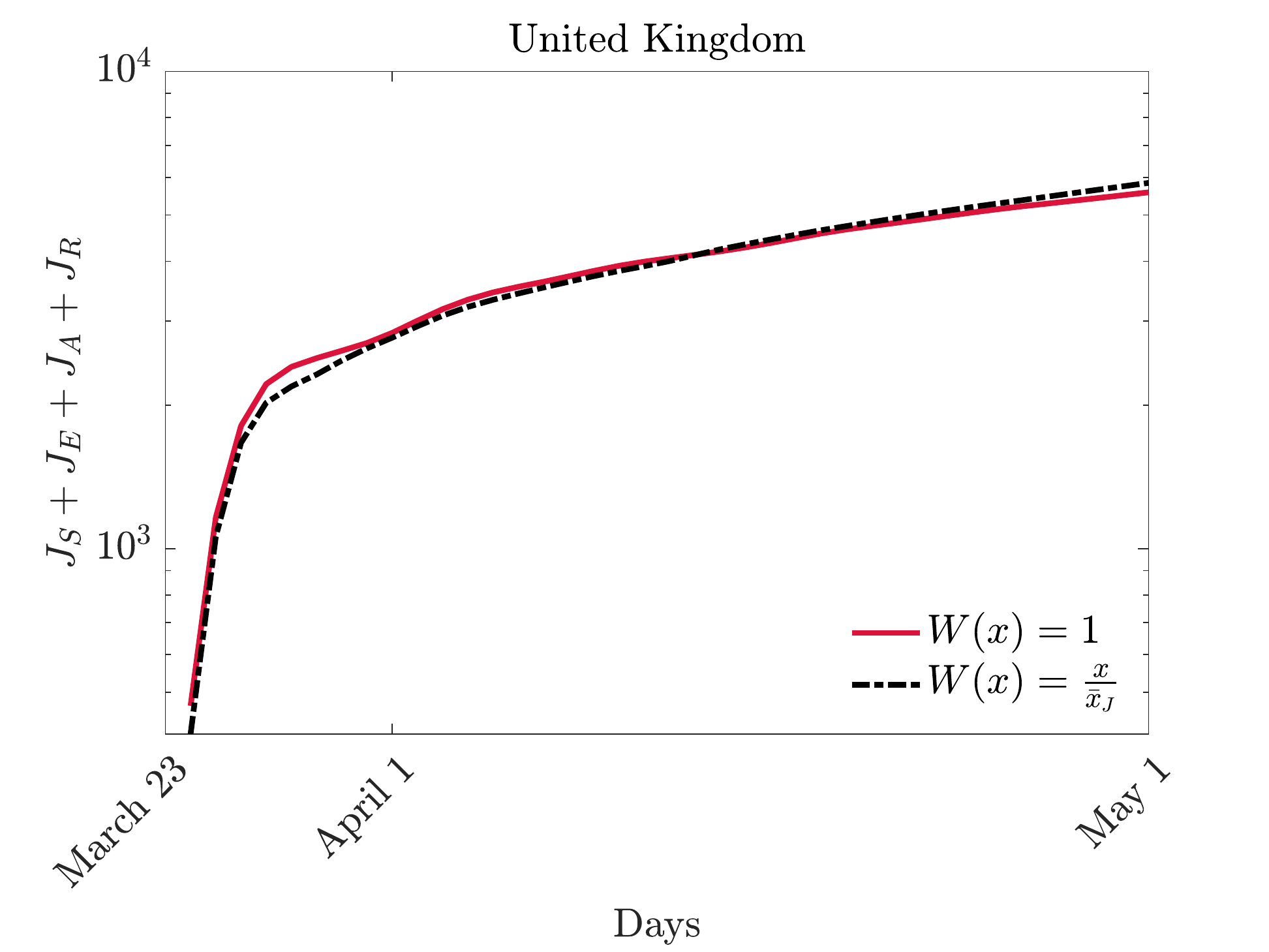}
\caption{Top: obtained fitting of the observed number of infected {(labelled with Data)} together with estimated number of exposed and asymptomatic cases. Middle: estimated target number of contacts $x_T$. Bottom: evolution of the sum of cost functionals $J_S + J_E + J_A + J_R$. In all the presented tests we compared the cases of uniform control ($W(x) = 1$) and selective control ($W(x)= x/\bar{x}_i$, $i\in \mathcal C$). }
\label{fig:good}
\end{figure}

Before proceeding with the estimation of the model parameters some remarks are deserved. First, we stress that the calibration of epidemiological models is generally difficult and require special cares at the numerical level. Furthermore, available data typically represent a lower bound with respect to the true numbers of an infection, especially in its initial phases, due to a limited testing capacity. Here, we do not consider this aspect and we refer to recent attempts to overcome this issue based on tools of uncertainty quantification \cite{APZ,Zanella} that can be successfully employed in this context. For related approaches we mention also \cite{Capaldi,BP,R,BBDP}. 

For the calibration of the model \eqref{seiar-mass}-\eqref{seiar-mean} we have adopted a bilevel approach . In particular, in the phase preceding the lockdown, we estimated the epidemic parameters in the unconstrained regime assuming that no restriction of social contacts  was activated,  i.e. $\mathcal C\equiv 0$. 
We also fixed the known clinical parameters according with values reported in the literature: see \cite{BDM,Gatto}. We set then $\gamma_E = 1/3.32$, $\gamma_I = 1/10$ and $\gamma_A = 2\gamma_I$, $\beta_A = 2\beta_I$ and we also assumed that $\xi = 0.2$, such that the $80\%$ of infections come from asymptomatic or paucisymptomatic individuals, in agreement with the recent serological campaigns promoted in Italy  \footnote{Preliminary results on the seroprevalence of SARS-CoV-2 in Italy:\\
\texttt{https://www.istat.it/it/files//2020/08/ReportPrimiRisultatiIndagineSiero.pdf}}. The proposed choice for $\beta_A,\beta_I>0$ is coherent with existing literature, see e.g. \cite{Parolini}. 

We then solved a least square problem based on the minimization of the relative $L^2$ norm of the difference between reported number of infected $\hat{N}_I$ and recovered $\hat{N}_R$, and the theoretical evolution of the model $N_I(t)$ and $N_R(t)$ for $t \in [t_0,t_\ell]$ with $N_E(t_0) = N_I(t_0) = N_A(t_0) = N_R(t_0) =  1$, i.e. we assume that in $t_0$ the pandemic was nearly started.
We also considered $\bar x_S(t_0) = \bar x_E(t_0) = \bar x_A(t_0) = \bar x_R(t_0) = 10$, in agreement with the experimentally observed mean number of contacts in a Western country before the pandemic \cite{Plos}, and $\bar x_I(t)\equiv 3$ for recognized infected, corresponding to the average number of family contacts. Thus once the known parameters have been fixed, we considered the following minimization problem
\[
\min_{\beta_I,\lambda} \left[(1-\theta)\|N_I(t)-\hat{N}_I(t) \|_{L^2([t_0,t_\ell])} + \theta\|N_R(t)-\hat{N}_R(t) \|_{L^2([t_0,t_\ell])}\right], 
\]
{where} $\theta \in [0,1]$, $\| \cdot \|_{L^2([t_0,t_\ell])}$ the relative norm over the time horizon $[t_0,t_\ell]$ and under the constraints $\beta_I \in [0,1/100]$, $\lambda \in [0,50]$. 
In third and fourth columns of Table \ref{tab:dates} we report the estimated epidemiological parameters obtained for $\theta = 10^{-3}$ to get a better resolution of the infected dynamics. It is worth to notice that, according to these values, Italy manifests higher heterogeneity linked to the contact distribution compared to UK. 

\subsection{Assessing the impact of non-pharmaceutical interventions}

Once the relevant epidemiological parameters are estimated, we focus on the subsequent lockdown phase where we look for the optimal value $x_{T}$ in the control term which permits to fit the data. We assume this value $x_T$ equal for each compartment and we study the two cases of uniform and selective restrictions, respectively obtained by fixing $W(x) = 1$ and $W(x) = x/\bar{x}_i$ and $\mathcal C(f^\infty)$ of the form \eqref{eq:Cinf} in the dynamics. Hence, we solve an optimization problem in the lockdown time interval $[t_\ell+1,t_f]$, for a sequence of time steps $t^n$  over a time window of one week
\[
\min_{x_T(t^n) \in \mathbb R^+} \left[(1-\theta)\|N_I(t)-\hat{N}_I(t) \|_{L^2([t^n - \kappa_\ell,t^n + \kappa_r])} + \theta\|N_R(t)-\hat{N}_R(t) \|_{L^2([t^n - \kappa_\ell,t^n + \kappa_r])}\right],
\]
with $\kappa_\ell = 3$, $\kappa_r = 4$.

\begin{figure}
\includegraphics[scale = 0.33]{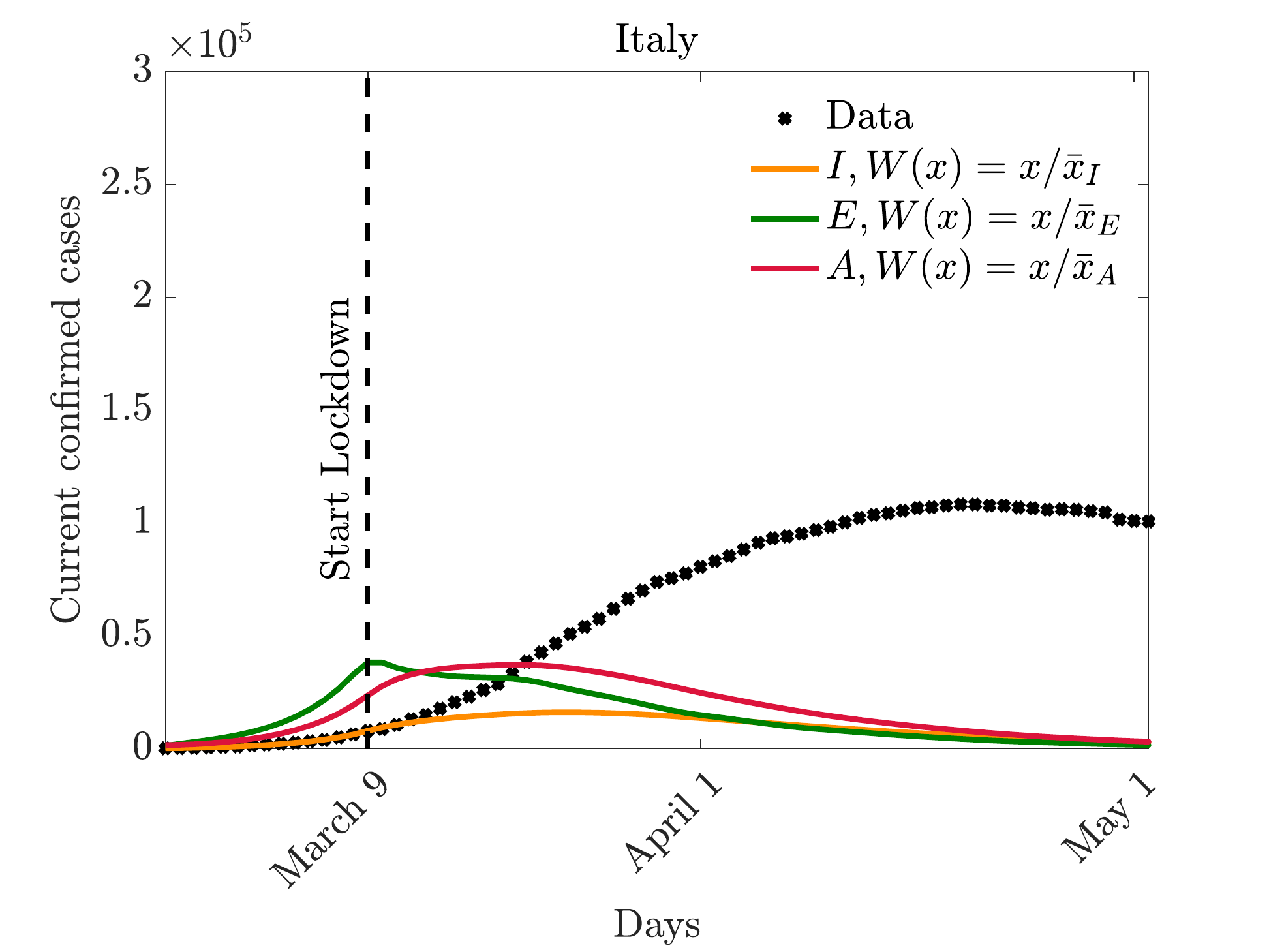}
\includegraphics[scale = 0.33]{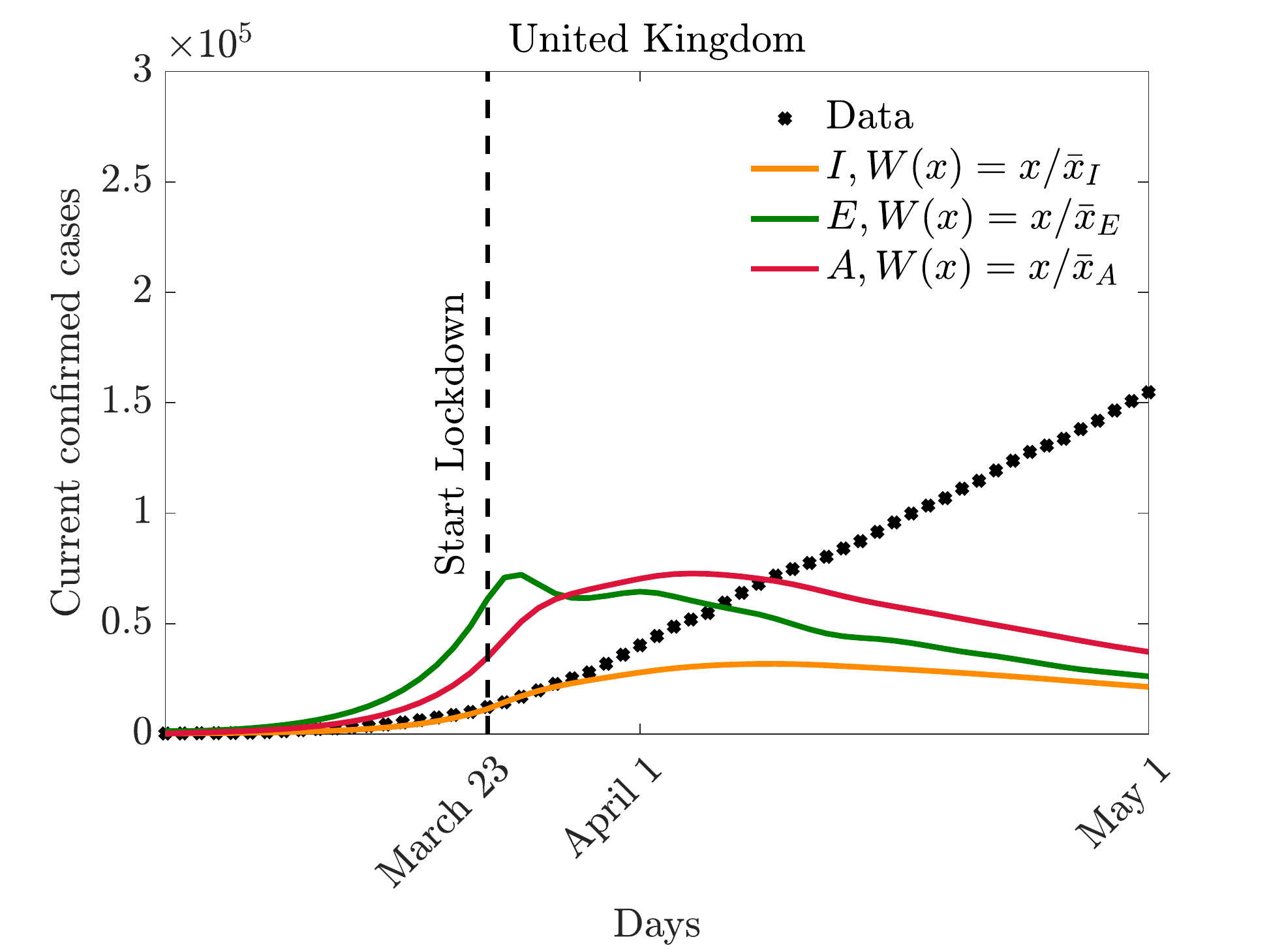}
\caption{Retrospective analysis of the epidemic dynamics in presence of selective control strategy with target $x_{T,i}$, $i \in \mathcal C$, estimated in the uniform case. }
\label{fig:xT}
\end{figure}

In Figure \ref{fig:good} we show the obtained trends for the registered cases for Italy and United Kingdom. We may observe how for both controls the macroscopic trends are coherent with the observed ones. Therefore, the adopted fitting procedure allows us to compare the estimated target $x_{T,i}$ for mean contacts in the considered control strategies. In particular, we observe how the uniform control implies a stronger contact reduction with respect to the selective one, while a similar cost, computed as the sum $ J_S + J_E + J_A + J_R$, is produced by the two containment strategies. 
In other words, the adoption of selective-type containment strategies based on the average number of contacts allows to mitigate the severity of lockdown policies.

In the previous part, we focused on getting the target mean number of contacts for both the control strategies. We remark  that the NPIs implemented by both countries have been mainly uniform in the first wave of the epidemic, meaning that, in the present model framework, the average number of contacts  where approximately the ones defined by the red line with circle markers in the second row of Figure \ref{fig:good}. Hence, it is interesting to consider a retrospective analysis where the estimated $x_{T,i}$, $i \in \mathcal C$, is instead implemented in the dynamics with selective control. In Figure \ref{fig:xT}, we present the results of such study, where the evolution of the disease in presence of selective control with target given by the uniform control is reported. It is evident how the amplitude of the epidemic peak is strongly reduced and the fraction of asymptomatic cases is under control in both cases using this alternative strategy. Furthermore, it is worth to observe that a selective-type control is more effective in reducing the peak of infection in presence of high heterogeneity like in the case of Italy. 

\section{Conclusion}
The recent spreading of COVID-19 epidemic has been thwarted in most countries through non-pharmaceutical interventions,   intended to diminish transmission rates and, to this end, reducing person-to-person contacts via social distancing. 
Even if the confinement measures appear to be a correct  tool to limit the infection spreading, a precise quantification of the effects played by them has still to be carefully understood. 

Aiming to investigate the impact of the number of daily contacts in the spread of infectious diseases, in \cite{DPTZ} we introduced a mathematical description of  the epidemic spreading by integrating the epidemiological dynamics of the classical SIR-type compartmental model  with a kinetic term quantifying the population-based contacts. 

The kinetic description of the formation of social contacts considered in \cite{DPTZ}, based on elementary interactions, 
turns out to be very flexible with respect to external inputs, like a control forcing the number of daily social contacts towards a fixed target. 
From the modeling point of view, this strategy leads to integrate the compartmental models with social contacts with a new term quantifying in a precise way the effect of the control strategy in dependence of the population heterogeneity. 

The main conclusion of our theoretical analysis, confirmed by numerical experiments, is that a selective control, when compared to a uniform control, and at the same social cost, leads to a marked reduction of the epidemic spreading. This reduction is further linked to the parameter of social heterogeneity, so that the selective control gives a better performance in a population with higher heterogeneity.

The main conclusion of our theoretical analysis, confirmed by numerical experiments, is that a selective control, when compared to a uniform control, and at the same social cost, leads to a marked reduction of the epidemic spreading. This reduction is further linked to the parameter of social heterogeneity, so that the selective control gives a better performance in a population with higher heterogeneity.

\section*{Acknowledgement} This work has been written within the
activities of GNFM and GNCS groups  of INdAM (National Institute of
High Mathematics). The research was partially supported by
the Italian Ministry of Education, University and Research (MIUR: Project ``Optimal mass
transportation, geometrical and functional inequalities with applications'', and
 Dipartimenti di Eccellenza Program (2018--2022) - Dept. of Mathematics ``F.
Casorati'', University of Pavia. 
G.D. would like to thank the Italian Ministry of Instruction, University and Research (MIUR) to support this research with funds coming from PRIN Project 2017 (No. 2017KKJP4X entitled ``Innovative numerical methods for evolutionary partial differential equations and applications'').


\end{document}